\documentstyle[aps,prd,12pt]{revtex}
\begin{document}
\tighten
\title{\bf Non-Archimedean character of quantum buoyancy\\
 and the generalized second law of thermodynamics} 

\author{Jacob D. Bekenstein\thanks{e--mail: bekenste@vms.huji.ac.il}
}

\address{\it The Racah Institute of Physics, Hebrew University
of Jerusalem,\\ Givat Ram, Jerusalem 91904, Israel}

\date{\today}

\maketitle

\begin{abstract}

Quantum buoyancy has been proposed as the mechanism protecting the
generalized second law when an entropy--bearing object is slowly lowered
towards a black hole and then dropped in.  We point out that the original
derivation of the buoyant force from a fluid picture of the acceleration
radiation is invalid unless the object is almost at the horizon, because
otherwise typical wavelengths in the radiation are larger than the
object.  The buoyant force is here calculated from the diffractive
scattering of waves off the object, and found to be weaker than in the
original theory.  As a consequence, the argument justifying the generalized
second law from buoyancy cannot be completed unless the optimal
drop point is next to the horizon.  The universal bound on entropy is always
a sufficient condition for operation of the generalized second law, and can
be derived from that law when the optimal drop point is close to the
horizon.  We also compute the quantum buoyancy of an elementary charged
particle; it turns out to  be negligible for energetic considerations. 
Finally, we speculate on the significance of the absence from the bound of
any mention of the number of particle species in nature.
\end{abstract} 

\section{Introduction}
\label{intro}
An observer accelerating in flat and empty spacetime with acceleration $a$
detects isotropic thermal radiation with temperature $T_U=\hbar a/2\pi$, the
celebrated Unruh radiance\cite{Unruh}.  An object suspended in the vicinity
of a black hole is accelerated by virtue of its being prevented from
following a geodesic.    Unruh and Wald (UW)\cite{UW1} suggested that
this object will likewise see Unruh radiance.  Since its acceleration (hence
$T_U$) varies with distance from the horizon, UW surmised that the object
will be subject to a buoyant force, just as an object submerged in fluid in a
gravitational field is buoyed up by the non--uniformity of the ambient
pressure.  Two intriguing consequences were inferred: the buoyancy can cause
an object sufficiently near the horizon to ``float'', and
the buoyancy affects the energetics of a process whereby an object is
lowered from afar towards a black hole while doing work on the lowering
mechanism\cite{UW1}.   

Both these effects bear on the important issue of what is the
mechanism enforcing the generalized second law (GSL)\cite{GSL} when an
entropy--bearing object is lowered slowly towards a black hole, and then
dropped in ?  Our initial inquiry\cite{Bound}, carried out before the
quantum buoyancy was put in evidence, assumed it is possible to lower the
object almost to the horizon, and inferred from the GSL that its entropy,
$S$, must be bounded in terms of its radius $R$ and mass--energy $E$ (units
with $c=1$ throughout) by
\begin{equation}
S\leq 2\pi RE/\hbar
\label{bound}
\end{equation}
This is to be regarded as a universal bound on the entropy of matter
(radiation) calculated to the deepest level of structure.  Bound
(\ref{bound}) is now backed by much independent evidence from flat space
physics (for a review see reference \cite{BekSchiff}), and was an antecedent
of 't Hooft's and Susskind's holographic
principle\cite{tHooft,Susskind}, which can be deduced from the bound
for weakly gravitating systems\cite{MG7,Bousso}.

UW questioned the black hole derivation of bound (\ref{bound}) on the grounds
that the quantum buoyancy allows free lowering of the object only down to
the floating point, and modifies the naive energetics of the problem.  
They developed a three--steps argument\cite{UW1} suggesting that buoyancy is
the only mechanism required for the GSL to work, and that bound (\ref{bound})
need not be invoked (and cannot thus be derived from the GSL).  The
steps are: 
\begin{itemize}
\item 1. Calculation of the full energy of the object at the floating point, 
an energy which determines the minimum increase in black hole entropy,
$\Delta S_{BH}$, occasioned when the object is dropped into the hole.  UW
claim that  $\Delta S_{BH}$ is never smaller than ${\cal S}_{\rm disp}$, the
entropy in the Unruh radiance displaced by the object at the floating point.  
\item 2. Demonstration that Archimedes' principle is obeyed: at the floating
point the energy of the displaced radiation equals the object's proper
energy.  
\item 3. Proposition that the object's entropy $S$ cannot possibly exceed
${\cal S}_{\rm disp}$ because unconfined thermal radiation is maximally
entropic for given volume and energy  (UW's entropy bound).  
\end{itemize}
Combination of the three steps gives $\Delta S_{BH}>S$: the GSL works
because the black hole entropy overcompensates for the loss of the object's
entropy.  Other arguments that quantum buoyancy is sufficient by itself to
protect the GSL have been offered by Li and Liu\cite{LiLiu},
Zaslavskii\cite{Zaslavskii} and  Pelath and Wald (PW)\cite{PW}.

Following UW's original paper we noted that, provided the object is
macroscopic (technically $\hbar/E\ll R$), and that the number $N$ of
species present in the Unruh radiance is moderate, the floating point is
necessarily very near the horizon\cite{Bound2,Bound3,newbound}.  As a
consequence the lowering process is almost unchanged by quantum buoyancy,
with buoyant corrections to its energetics showing up late in the lowering
process.  This has allowed bound (\ref{bound}) to be rederived from the GSL
applied to the described gedanken experiment even in the face
of buoyancy\cite{LiLiu,newbound,Zaslavskii}.   

PW\cite{PW} have kindled the controversy anew by asserting that this
feat is made possible only because the same assumptions whose outcome is to
put the floating point very near the horizon will cause bound (\ref{bound}) to
be satisfied automatically without any recourse to black hole physics.  PW
base their argument on UW's bound on entropy (step 3 above). Exceptions to
the UW's bound have been demonstrated in flat spacetime\cite{Bound3}: an
object's entropy can exceed that of an equal energy and volume of unconfined
thermal radiation if one (or two) of the object's dimensions is (are) much
larger than the third.  In the present paper we mostly focus on a spherical
object for which the UW bound may well hold; this allows us to shift
attention to the first two steps of UW's argument. 

So is bound (\ref{bound}) a condition for the GSL to work in accretion by a
black hole ?  Or is quantum buoyancy the sole protector of that law, so that
the bound cannot be inferred from it ?   Given the importance of the GSL as
an extension of the very pervasive second law, and as giving operational
meaning to the concept of black hole entropy, it is obviously important to
elucidate the questions raised.  In addition, one would like to clarify
how allowing for the existence of an arbitrary number of species in the
radiation bears on all earlier conclusions.  Both UW and PW are of
the opinion that the GSL should be shown to be upheld by a single mechanism
for arbitrary $N$.  The present paper provides substantial clarification of
all these issues.

In Sec.\ref{new} we point out that buoyancy away from the horizon has
been calculated incorrectly in all extant works.  Departures from the fluid
description of the Unruh radiance result because wavelenghts in the radiance
are typically longer than the object's size.  Sec.\ref{far} deals with
a spherical object always removed from the horizon by a proper distance
larger than its own size.  In Sec.\ref{momentumcurved} we calculate the
buoyancy on it, and find it to be weaker and more rapidly falling with
distance from the horizon than would be expected from UW's
approach.  Sec.\ref{energeticsFR} we make it clear that Archimedes' principle
fails in this far region, and also that UW's  ``GSL from buoyancy'' argument
cannot be completed.  By contrast, the entropy bound (\ref{bound}) is found
to be a sufficient condition for the GSL to function. 

In Sec.\ref{near} we consider an object whose floating point is much
closer to the horizon than its own size.  This case is relevant for a
macroscopic object and a moderate number of radiance species.  UW's fluid
estimate of the buoyancy force is correct in this near region; however,
their energetics are slightly changed because the object has to pass through
the far region on its way down.  Overall the UW argument that the GSL is
protected by the buoyancy is upheld in the near region.  On the other hand,
in spite of the buoyancy, the entropy bound (\ref{bound}) is derivable from
the GSL.  In Sec.\ref{intermediate} we treat the intermediate region where
the object's proper distance from the horizon is comparable to its size, and
confirm the conclusion that the argument for the GSL from buoyancy cannot
be completed over a certain regime; the entropy bound remains a sufficient
condition for the GSL to work.

In Sec.\ref{elementary} we consider buoyancy of an elementary charged
particle, finding it to be negligible under all circumstances.  This is
important for one of the arguments supporting the uniformly spaced black hole
area quantum spectrum\cite{MG8,Brazil,Hod} 

\section{Fluid vs Wave picture}
\label{new}

All mentioned works follow UW in assuming that the object is impervious to 
the radiation, and that the buoyant force, as measured at infinity, can be
calculated by integrating the radiance pressure $p$ multiplied by the local
redshift factor $\chi$ and by the inward normal to the surface all around
the object's surface.  UW\cite{UW1} relied on thermodynamic description of
the radiance as a fluid obeying  $e+p-Ts=0$,  where $e$ denotes the proper
energy density, $s$ the proper entropy density, and $T$ the local
temperature. They took $T$ in the vicinity of the horizon to obey
\begin{equation}
T= T_0\,\chi^{-1},
\label{T_0}
\end{equation}
where $T_0$ denotes the Hawking temperature.  One can justify this form by
noting it is a good approximation to the formal Unruh temperature
corresponding to the acceleration felt by a point suspended at rest in a
Schwarzschild metric at distances from the horizon small compared to the
size of the black hole\cite{newbound,Zaslavskii}. 

While there is little doubt about the cogency of the thermodynamic
description of the {\it properties\/} of the radiation, it seems to have gone
unnoticed in the past that the conditions encountered by a small object near
the horizon may make the calculation of the buoyant force from
the fluid point of view inappropriate.  Let $R$ denote the typical size of
said object; we shall assume $R\ll GM$ ($M$ is the hole's mass) in order that
the full object may be able to approach the horizon.  For a Schwarzschild
black hole's exterior the exact metric is (units with $c=1$)
\begin{equation}
ds^2 = -\chi^2 dt^2 + dz^2 + r^2(d\theta^2 + \sin^2 \theta d\phi^2)
\label{metric}
\end{equation}
where $z=\int_{2GM}^r (1-2GM/r)^{-1/2} dr$ is the proper radial distance from
the horizon while $\chi=(1-2GM/r)^{1/2}$ is the redshift factor mentioned
earlier.  Our calculations will be done in the region $r-2GM\ll GM$
(equivalently $z\ll GM$); it is only in it that we may approximate
Unruh's temperature by Eq.(\ref{T_0}), and at the same time ignore Hawking's
radiance which is not manifest so near the horizon\cite{UW1}.  We then get 
\begin{equation}
\chi\approx z(4GM)^{-1}.
\label{chi}
\end{equation}
In the Schwarzschild case $T_0=\hbar(8\pi
GM)^{-1}$, so we have\cite{newbound,Zaslavskii} 
\begin{equation}
T \approx \hbar (2\pi z)^{-1}.
\end{equation}

Thus we expect that the predominant local wavelength in the radiation,
$\bar\lambda$, at proper distance $z$ from the horizon is roughly of order 
$z$.  In fact, distribution (\ref{Planckian}) to be obtained below peaks a
$\lambda\approx 7.97 z$ and gives the mean wavelength as
$\langle\lambda\rangle = 14.62 z$.  So if the object is near the horizon
 ($z\ll GM$) but not nearly touching it
($z{\,\,\lower0.5ex\hbox{$\sim$}\kern-10pt\raise0.5ex\hbox{$>$}\,\,} R$), we
have
$R{\,\,\lower0.5ex\hbox{$\sim$}\kern-10pt\raise0.5ex\hbox{$<$}\,\,}
z\sim\bar\lambda\ll GM$.  At a fundamental level the buoyant force is due
to the momentum jolts the object receives as successive waves scatter off it.  
Waves with
$\lambda{\,\,\lower0.5ex\hbox{$\sim$}\kern-10pt\raise0.5ex\hbox{$>$}\,\,} R$
have difficulty matching specified boundary conditions on the object's
surface; hence they tend to scatter poorly and convey little momentum to the
object.  Waves with
$\lambda\ll R$ can match the boundary conditions better, and scatter and
convey momentum effectively.  However, since these short waves are a minority
in the Unruh radiation ($\lambda\ll
R{\,\,\lower0.5ex\hbox{$\sim$}\kern-10pt\raise0.5ex\hbox{$<$}\,\,}
\bar\lambda$), we expect the true buoyant force to be small compared to that
we would calculate by means of the radiation pressure (which comes from all
wavelengths indiscriminately); in other words, the buoyancy should be
non--Archimedean.  Detailed calculation in Sec.\ref{momentumcurved} confirms
this expectation.  Since the object being lowered must pass through the
region with
$z{\,\,\lower0.5ex\hbox{$\sim$}\kern-10pt\raise0.5ex\hbox{$>$}\,\,} R$ - we
call it the {\it far region\/} or FR - the energy accreted by the black
hole and its entropy increase when the object is finally dropped in must be
influenced by the suppression of buoyancy just mentioned.  One purpose of
this paper is to correct UW's treatment of the lowering process for the
suppression, and to reexamine quantum buoyancy's role in the GSL's operation.

\section{Buoyancy in far region}
\label{far}

We first concentrate on the situation where the object's bottom is at no
time closer (in proper distance) to the horizon than the object's height. 
The opposite situation will be treated in Sec.\ref{near}.

\subsection{Momentum transfer in flat spacetime}
\label{momentumflat} 

Obviously, a force calculation from wave momentum transfer is bound to be
complex as compared to one based on the fluid picture, so we start by
discussing the scattering in flat spacetime.  It is well known that wave
scattering by an object in the long wavelength limit is indifferent to
details (target's shape, {\it etc.\/}).  The differential crossection has the
form
\begin{equation}
d\sigma/d\Omega = R^2(R/\lambda)^4F(\hat {\bf n}, \hat {\bf n}')
\label{crossection}
\end{equation} 
where $\pi R^2$ is a typical geometric crossection of the object, $F$
some dimensionless function, and
$\hat {\bf n}$ and
$\hat {\bf n}'$ a pair of unit vectors denoting the incidence and scattering
directions, respectively.  The fourth order dependence on wavelength comes
from the dipole part of the scattering, which predominates at long
wavelengths.  As an example\cite{Jackson}, electromagnetic scattering from a
conducting sphere has
$F(\hat {\bf n}, \hat {\bf n}') = 16\pi^2[{\scriptstyle5\over
\scriptstyle 8}(1+\cos^2 (\hat{\bf n}\cdot\hat{\bf n}'))-\cos (\hat{\bf
n}\cdot\hat{\bf n}')]$.  If the incident flux of momentum carried in the
ambient radiation by wavelengths in the vicinity of $\lambda$ and in the
vicinity of the direction $\hat{\bf n}$ is $\hat{\bf n}\,I(\lambda, \hat {\bf
n})\,d\lambda\, d\hat {\bf n}$, the object will gain momentum at a rate 
\begin{equation}
d{\bf P}/dt=\int d\lambda \int d\hat{\bf n}\,\hat {\bf n}\,I(\lambda, \hat
{\bf n})R^2(R/\lambda)^4\,\int d\hat{\bf n}' F(\hat{\bf n}, \hat{\bf
n}')\,(1-\hat{\bf n}\cdot\hat{\bf n}').
\label{dP}
\end{equation}
The factor $1-\hat{\bf n}\cdot\hat{\bf n}'$ takes into account extra momentum
given to the object when a wave backscatters.  

Because characteristics of the scattering are insensitive to the object's
shape, we may narrow attention to a spherically symmetric object.  Results for
objects not too far from spherical should be quite similar.  In what
follows   $R$ shall denote the object's radius. The scattering crossection
now depends only on the angle of scattering: $F=F(\hat{\bf n}\cdot\hat{\bf
n}')$.  The integral over the solid angle $\hat{\bf n}'$ is then equivalent
to $2\pi$ times one over
$\hat{\bf n}\cdot\hat{\bf n}'$; it is thus just a number $\xi$, most likely
of order unity (for electromagnetic waves $\xi=14\pi/3$).  Thus
\begin{equation}
d{\bf P}/dt=\xi \int d\lambda \int d\hat{\bf n}\,\hat{\bf n}\,I(\lambda,
\hat {\bf n})R^2(R/\lambda)^4.
\label{scat}
\end{equation}

\subsection{Buoyant force in curved spacetime}
\label{momentumcurved}

In curved spacetime, particularly in the vicinity of a black hole, we may
take over Eq.(\ref{scat}) to an orthonormal frame associated with metric
(\ref{metric}) in which the object is at rest.  As usual, many quantities, 
such as field strengths, fluxes, wavelengths, {\it etc.} will take on the
same values in the orthonormal frame and in a comoving inertial frame; such
inertial frame can be large enough to contain the object and its immediate
surroundings because  by assumption the radius of curvature at the horizon, 
$GM$, is much larger than $R$. This allows us to use flat spacetime results
about the scattering.  We should reinterpreted  ${\bf P}$, $\lambda$ and $t$
above as momentum, wavelength and time measured in the orthonormal frame. 
And this time must be proper time $\tau$ of the object, while $d{\bf
P}/d\tau$ stands for the force as measured by a local observer in the
orthonormal frame.  With this in mind we may rewrite Eq.(\ref{scat}) as
\begin{equation}
d{\bf P}/d\tau=\xi R^6  \int d\lambda \int d\hat{\bf n}\,\hat{\bf
n}\,I(\lambda,
\hat {\bf n})\,\lambda^{-4}.
\label{dp/dt}
\end{equation}

Since the region including the object and its immediate surroundings is
small on scale $GM$, and close to the horizon, $r\approx 2GM$ there.  It
proves more convenient to rewrite the metric (\ref{metric}) in said region in
the approximate form
\begin{equation}
ds^2 = -(z/4GM)^2 dt^2 + dz^2 +dx^2 +dy^2
\label{metric2}
\end{equation}
where $dx^2 +dy^2$ is the metric on the almost planar small cap of the sphere
$r=$ const.$\approx 2GM$ which has fixed $z$.  The directions $\hat{\bf n}$
and $\hat{\bf n}'$ have the same components in appropriately
oriented inertial and orthonormal frames as with
respect to the spacelike coordinate lines of metric (\ref{metric2}).

We proceed to calculate $I(\lambda, \hat {\bf n})$ starting in the
global frame defined by Eq.(\ref{metric2}).  Because that metric is
static, the global frequency $\omega_0$ of a wave is conserved (being the
same over the full black hole exterior described by metric (\ref{metric})). 
And because the system is described by a single global temperature, $T_0$,
one can describe the thermal spectrum by saying that each global mode of a
boson field with frequency $\omega_0$ is occupied by
$[\exp(\hbar\omega_0/T_0)-1]^{-1}$ quanta on average.  To find the flux of
momentum at a given point we have to enumerate all contributing modes in a
small region surrounding the spherical object.  To make things as simple as
possible consider the scalar equation
$\Phi_{,\alpha}{}^{;\alpha}=0$ on the metric (\ref{metric2}):
\begin{equation}
\Phi_{,uu} + (z^*/4GM)^2 \exp(u/2GM)(\Phi_{,xx}+\Phi_{,yy}) - \Phi_{,tt}=0
\end{equation}
Here $u\equiv 4GM \ln(z/z^*)$, and $z^*$ is a reference value of the $z$
coordinate.  The symmetries in $x$, $y$ and $t$ permit a solution of the form
\begin{eqnarray}
\Phi&=&U(u) \exp[\imath(\kappa_x x+ \kappa_y y -\omega_0 t)];
\label{mode}
\\
0&=&U_{,uu} + [\omega_0^2 - (z^*/4GM)^2 \exp(u/2GM) (\kappa_x^2+\kappa_y^2)]U
\label{modeequation}
\end{eqnarray}
where $\kappa_x$, $\kappa_y$ and $\omega_0$ are constants, the last
identical to the global frequency mentioned above.
 
For given $\kappa_x$ and $\kappa_y$ there are oscillatory solutions of
Eq.(\ref{modeequation}) for sufficiently large $\omega_0$.  In WKB
approximation they take the form
\begin{eqnarray}
U&\approx& \exp\left\{\imath\int[\omega_0^2 - (z^*/4GM)^2 \exp(u/2GM)
(\kappa_x^2+\kappa_y^2)]^{1/2} du\right\}
\\
&=&\exp\left\{\imath\int[\omega_0^2 (4GM/z)^2 - 
(\kappa_x^2+\kappa_y^2)]^{1/2} dz\right\}
\end{eqnarray}
From the last form, which is $z^*$ independent, we infer the effective
wavevector component in the $z$ direction: $\kappa_z = [\omega_0^2 (4GM/z)^2
- (\kappa_x^2+\kappa_y^2)]^{1/2}$.  Although the accelerated character of
the global coordinates causes this component to vary, over the
short $z$ interval encompassed by the sphere we can think of
$\{\kappa_x,\kappa_y,\kappa_z\}$ as a wavevector.  Obviously the
corresponding wavelength is $\lambda
=2\pi(\kappa_x^2+\kappa_y^2+\kappa_z^2)^{-1/2}=\pi z(2GM\omega_0)^{-1}$,
which is just what we would have gotten had we corrected the global frequency
$\omega_0$ for the redshift $\chi$.  $\lambda$ thus coincides with the 
locally measured wavelength we have been using intuitively.  

Now according to Eq.(\ref{metric2}), $x$, $y$ and $z$ measure proper length;
therefore, per unit proper volume, the number of allowed values  of
$\kappa_x$, $\kappa_y$ and $\kappa_z$ in an interval $d  \kappa_x$, $d 
\kappa_y$ and $d \kappa_z$ is given by $(2\pi)^{-3}d  \kappa_x d  \kappa_yd
\kappa_z$, or equivalently, by
$\lambda^{-4}\,d\hat{\bf n}\,d\lambda$, where $d\hat{\bf n}$ is the solid
angle spanned by the various wavevectors. Taking two helicities for each
wavevector (here we pass from scalar waves to electromagnetic), and
remembering that the momentum carried by a singly occupied mode, as measured
in the orthonormal frame, is $2\pi\hbar\lambda^{-1}$, we obtain the momentum
flux density $I(\lambda, \hat {\bf n})$ in the orthonormal frame,
\begin{equation}
I(\lambda, \hat {\bf n}) = {4\pi\hbar\lambda^{-5} \over e^{2\pi\hbar\over
\lambda T}-1}
\label{Planckian}
\end{equation}
Here we have used Eq.(\ref{T_0}) to replace $\hbar\omega_0/T_0$ by
$2\pi\hbar(\lambda T)^{-1}$. 

We should not be fooled by the similarity of this expression to the Planckian
distribution in textbooks.  Because of the variation of the redshift $\chi$,
$\lambda$ of a particular mode changes (redshifts) as the wave propagates
(and thus passes from one orthonormal frame to a neighboring one). 
Alternatively, if we ask how much momentum is carried by the radiation in
$(\lambda,\lambda+d\lambda)$, we must be aware that $T$ in Eq.(\ref{Planckian})
varies with position according to Eq.(\ref{T_0}).  This has implications for
the integral over $\hat{\bf n}$ in Eq.(\ref{dp/dt}).   The temperature $T$ to
be used therein is not isotropic because it is not meant to be taken at a
fixed point.   A (plane) wavefront incident on the sphere with direction
$\hat{\bf n}$ first meets its surface at coordinate $z=z_c-\hat{\bf z}\cdot
\hat{\bf n}\, R$, ($\hat{\bf z}$ is a unit vector in the positive  $z$
direction and $z_c$ is the sphere center's $z$ coordinate).  Therefore,
it is reasonable to evaluate the temperature governing the
incident radiance's intensity at this $z$.   In view of Eq.(\ref{T_0}), we
must take 
\begin{equation}
T=(\hbar/2\pi) (z_c-\hat{\bf z}\cdot \hat{\bf n}\, R)^{-1}
\label{anis}
\end{equation} in
Eq.(\ref{Planckian}) when evaluating the integral in Eq.(\ref{dp/dt}). 

We do the $\lambda$ integral first.  Of course, the dependence $\lambda^{-4}$
assumed for the crossection is not valid for $\lambda
{\,\,\lower0.5ex\hbox{$\sim$}\kern-10pt\raise0.5ex\hbox{$<$}\,\,} R$.  But
according to Eq.(\ref{T_0}), the exponential appearing in
Eq.(\ref{Planckian}), $\exp(4\pi^2 z/\lambda)$, is large when $\lambda < R$
because by assumption $R<z$: the Planckian distribution suppresses
contributions to $d{\bf P}/d\tau$ from short wavelengths.  Thus we may extend
the integration range down to $\lambda=0$ while incurring only a small error. 
After rescaling and using the identity
$\int_0^\infty x^7(e^x-1)^{-1}dx=8\pi^8/15$ we get
\begin{equation}
{d{\bf P}\over d\tau}={\xi\, \hbar R^6\over 30720\,\pi^{7}}
\int {\hat{\bf n}\,d\hat{\bf n}\over (z_c-\hat{\bf z}\cdot \hat{\bf n}\,
R)^8}
\label{buoy2}
\end{equation} 

By symmetry the integral must be proportional to $\hat{\bf z}$; it is
easily evaluated by going over to the variable $\hat{\bf z}\cdot \hat{\bf n}$.
To get the total FR buoyant force as measured at infinity, ${\bf f}_>$, one
must multiply $d{\bf P}/d\tau$, the buoyant force measured locally, by
$d\tau/dt$ of the sphere, which equals $\chi\approx z_c(4GM)^{-1}$.  We also
multiply by $N$, the effective number of species of quanta in the
radiation, assuming the object scatters all (photons contribute unity;
because radiations of different kinds scatter differently, 
$N$ is not necessarily an exact integer).  Thus, 
\begin{equation}
{\bf f}_>={\hat{\bf z}\,N\,\xi\, \hbar R^7\over
11520\,\pi^{6}\,GM}\,{z_c^6+2 R^2 z_c^4+3 R^4 z_c^2/7\over
\left(z_c^2-R^2\right)^{7}}.
\label{farforce}
\end{equation} 
Although we have carried out all integrals exactly, it must be understood 
that Eq.(\ref{farforce}) may already become inaccurate for
$z_c{\,\,\lower0.5ex\hbox{$\sim$}\kern-10pt\raise0.5ex\hbox{$<$}\,\,} 2R$
because in that region the condition $\lambda \gg R$ is satisfied only
marginally (as mentioned we expect $\bar\lambda\sim 10 z_c$).  The region $z_c
{\,\,\lower0.5ex\hbox{$\sim$}\kern-10pt\raise0.5ex\hbox{$<$}\,\,} 2R$ will be
examined in Sec.\ref{near}.  

We have just found that for $z_c\gg R$, ${\bf f}_> \propto z_c^{-8}$.  By
contrast, the force calculated {\it a la\/} UW's fluid approach,
Eq.(\ref{nearforce}) below, would behave in that region as
$z_c^{-4}$.  The corrections to the fluid approach's results are thus, {\it a
priori\/}, nontrivial in the FR.  

\subsection{Energetics in far region and GSL}
\label{energeticsFR}

Following UW we calculate the energy of the object as it is  slowly lowered 
from infinity.  As in many previous
papers\cite{UW1,Bound,PW,Bound2,Bound3,newbound}, we assume that the
object's proper energy $E$ and its entropy $S$ are unaffected by the descent.
The object's  gravitational energy at any stage is $E_{\rm grav}=E\chi
\approx E z_c(4GM)^{-1}$.   The contribution to the energy from the work done
to overcome the buoyancy is $\int^\infty_{z_c} \hat{\bf z} \cdot {\bf f}_>\,
dz_c$.  Of course our expression (\ref{farforce}) is not reliable for $z_c
{\,\,\lower0.5ex\hbox{$\sim$}\kern-10pt\raise0.5ex\hbox{$>$}\,\,} GM$;
however, the buoyant force drops off so fast with $z_c$ that we make a small
error if w use Eq.(\ref{farforce}) all the way to infinity.  Doing the
integral exactly we find the total energy function
\begin{equation}
{\cal E}_>(z_c)={E z_c\over 4GM} + {\xi\,N\,\hbar\,R^7
z_c^3\,(z_c^2+R^2)\over 80640\, \pi^6\, GM (z_c^2-R^2)^6}.
\label{E>}
\end{equation}

As UW point out, the most stringent test of the GSL is had by dropping the
object from the floating point where ${\cal E}_>$ reaches its
minimum.  This point is situated at $z_c=\ell R$  with $\ell$ the positive
root of
\begin{eqnarray}
{(\ell^2-1)^7\over \ell^6+2\ell^4+3\ell^2/7} = {\xi\sigma^2 \over 16 \pi^5}
\label{root}
\\
\sigma \equiv (N\hbar/180\pi ER)^{1/2}
\label{sigma}
\end{eqnarray}
We shall find $\sigma$ to be the main parameter delimiting various
buoyancy regimes.  The following flat spacetime intuition of it is useful. 
Consider an object of radius $R$ and energy $E$ immersed in thermal
radiation comprising $N$ species whose energy density equals the object's
energy density $\sim E/R^3$.  The usual Boltzmann formulae allow us to
conclude that the typical wavelength $\bar \lambda$ is of order
$\surd\sigma R$.  Thus a fluid (continuum) description of the radiation's
interaction with the object will be a good one for $\sigma\ll1$.  If the
object is macroscopic or even mesoscopic, its Compton length $\hbar/E$ must
be much smaller than $R$, so the requirement on $\sigma$ will hold nicely
provided $N$ is not large.
 
In the black hole spacetime, for the floating point to be in the FR
means $\ell {\,\,\lower0.5ex\hbox{$\sim$}\kern-10pt\raise0.5ex\hbox{$>$}\,\,} 
2$.
According to Eq.(\ref{root}) this requires  $\sigma
{\,\,\lower0.5ex\hbox{$\sim$}\kern-10pt\raise0.5ex\hbox{$>$}\,\,} 10^2$,
definitely not the fluid regime as we have realized already.   This large a
$\sigma$ requires $N$ of the Unruh radiance to be very large ($10^6$ at
least).  Although we are only aware of relatively few particle species in 
nature, it is of interest, as stressed by UW, to check whether the GSL would 
continue to work in the face of a proliferation of radiance species.

UW propose (step 2 in Sec.I here) that the floating point occurs precisely
where $E$ equals the proper energy of the displaced radiation $\int e\,
dx\,dy\,dz$.  The energy density of radiation
corresponding to spectrum (\ref{Planckian}) has the textbook form $e=(N\pi^2
T^4/15 \hbar^3)$.  We substitute Eq.(\ref{T_0}) and integrate over a
spherical volume equal to the sphere's and at the same height above the
horizon.  Since the horizontal crossection of the sphere at height
$z$ is $\pi[R^2-(z_c-z)^2]$, the integral becomes
\begin{equation}
{N\hbar\over 240\pi^2}\int_{z_c-R}^{z_c+R}\,{\pi[R^2-(z-z_c)^2]\over z^4}\,dz
= {N\hbar R^3\over 180\pi(z_c^2-R^2)^2}
\label{int_e}
\end{equation}
This equals $E$ at $z_c=R\sqrt{1+\sigma}$ which is distinct from $R\ell$:
contrary to UW's step 2, here the equal--energies point is
distinct from the floating point, and Archimedes' principle is invalid. 
Thus the buoyancy in the FR may be termed non--Archimedean.

This setback does not by itself disable the UW program.  The overall
argument might still work if one could prove that the minimum $\Delta S_{BH}$
exceeds the entropy contained in a spherical volume of Unruh radiance equal to
the object's and centered at $z_c=R\sqrt{1+\sigma}$ where it has the
same proper energy as the object.  We call this last the displaced entropy,
${\cal S}_{\rm disp}$, and evaluate it by integrating $s=(4e/3T)$ over said
spherical volume.  In terms of 
$h(x)\equiv \ln(\sqrt{1+x}+1)-\ln(\sqrt{1+x}-1)$,
\begin{equation}
{\cal S}_{\rm disp} =
(N/90)\big[2\sigma^{-1}\sqrt{1+\sigma}-h(\sigma)\big]
\label{Sdisp}
\end{equation}

Now in terms of $\ell$ and $\sigma^2$, ${\cal E}_>(\ell R)$ is easily
rewritten as
\begin{equation}
{\cal E}_>(\ell R)={E R\over 4GM}\left[\ell + {\xi\sigma^2\over
112\pi^5}{\ell^3(\ell^2+1)\over (\ell^2-1)^6}\right]
\label{E_<}
\end{equation}
Expressing $ER$ by means of Eq.(\ref{sigma}) and $\xi\sigma^2$
inside the square brackets by means of Eq.(\ref{root}),  and dividing by
$T_0$ gives the minimum increase in black hole entropy:
\begin{equation}
\Delta S_{BH} ={N\over 90\sigma^2}\,{\ell(8\ell^4+14\ell^2+2)\over
7\ell^4+14\ell^2+3}
\label{DSBH}
\end{equation}
A plot of both $\Delta S_{BH}$ and ${\cal S}_{\rm disp}$ as functions of
$\ell$ with $\sigma$ determined by Eq.(\ref{root}) shows that actually
$S_{BH}<{\cal S}_{\rm disp}$ already for $\ell>1.24$ (we took $\xi=10$ but
the results change little over a large range of $\xi$).  Thus the
inequality required by UW is violated over the whole FR where buoyancy has
to be treated as due to wave scattering.   

This being the case, the only way one could establish the GSL's validity
from the effect of the buoyancy would be to show that ${\cal S}_{\rm disp}$
is not just as large as $S$, but sufficiently larger so as to
compensate for the failure of $\Delta S_{BH}>{\cal S}_{\rm disp}$ whenever 
$\sigma$ is not small.  This
can no longer follow from a general principle such as UW's entropy bound,
but would entail looking into details of the object, something
quite opposite to UW's intent to establish the GSL in general.  We conclude
that in the FR UW's argument that buoyancy automatically enforces the GSL for
any $N$ {\it cannot\/} be completed.

On the other hand, if we substitute $\sigma^2$ from Eq.(\ref{sigma}) into
Eq.(\ref{DSBH}), and observe that the function of $\ell$ appearing in this
last one exceeds unity for $\ell>1$, we conclude that
\begin{equation}
\Delta S_{BH} > 2\pi ER/\hbar 
\end{equation}
It follows from bound (\ref{bound}) that the GSL is satisfied as the object
falls into the hole: the decrease of entropy by $S$ is compensated by  the
gain $\Delta S_{BH}$.  Bound (\ref{bound}) is thus a {\it sufficient\/}
condition for the operation of the GSL when the floating point is in the
FR.  
 
\section{Buoyancy in near--horizon region}
\label{near}

We now look at the situation where the object's bottom eventually comes
much closer to the horizon than the object's radius, so that it is in
the {\it near--horizon region\/} (NHR).  Because $\bar\lambda$ for waves
hitting the bottom is of order 10($z_c-R)$, and we shall want to keep it 
small compared to $R$, it is convenient to define the NHR as
$R<z_c{\,\,\lower0.5ex\hbox{$\sim$}\kern-10pt\raise0.5ex\hbox{$<$}\,\,}1.01R$.

\subsection{The buoyant force}
\label{buoyant} 

With the sphere so positioned the wavelengths of
waves hitting a cap at the sphere's bottom ($\lambda\sim z_c-R$) are short
compared to the cap's transversal size $\sim R$.  In this case the fluid
model can be used to compute the force on the cap.  As we go up the sphere's
surface, the waves hitting it get longer and eventually are no longer short
compared to the scale of the surface.  Somewhere along the side of the sphere
the fluid model fails.  Simultaneously, as we go up the surface, the
radiation (fluid) pressure 
\begin{equation}
p={N\pi^2 T^4\over 45\hbar^3}={N \hbar\over 720\pi^2 z^4}
\label{pressure}
\end{equation}
drops precipitously as $z$ grows.  Thus if we formally integrate the
pressure force all around the sphere, the main contribution comes from the
cap of about a steradian in size at the bottom which is almost horizontal and
very close to the horizon everywhere.  And this is precisely the part of the
force which is well described by the fluid model.   The force should be
corrected for the contribution from the rest of the sphere which is in the
wave scattering regime.  However, we know that the wave scattering force
tends to be weaker than the fluid force.  Hence, our integral must give a
close approximation to the true force.  The preceding comments amount to a
justification, when $z_c-R\ll R$, of UW's method for calculating the buoyant
force.

As UW mention, the pressure has to be multiplied by $\chi$ before
integration, so that the force will be ``as measured at infinity''. 
Therefore, working in metric (\ref{metric2}), the force in the NHR is
\begin{equation}
{\bf f}_< = -\oint \chi\, p\, d{\bf S} = -\int \nabla\,(\chi\, p)
dx\,dy\,dz
\end{equation}
We have used Gauss' theorem to convert the integral into one over the
volume the radiation would have occupied but for the sphere's presence.  As
in Eq.(\ref{int_e}) we rewrite this as
\begin{equation}
 -\int_{z_c-R}^{z_c+R}\,\nabla\,(\chi\,
p)\,\pi[R^2-(z-z_c)^2]\,dz = -2\pi \hat {\bf z} \int_{z_c-R}^{z_c+R}
\chi\, p\, (z-z_c)\, dz
\end{equation}
where we have done an integration by parts.  Substituting
Eqs.(\ref{pressure})  and (\ref{chi}) gives
\begin{equation}
{\bf f}_< = {\hat{\bf z} N\hbar\over 720\pi GM}\,{R^3\over
(z_c^2-R^2)^2  }
\label{nearforce} 
\end{equation}

\subsection{Energetics in near--horizon region and GSL}
\label{energeticsNHR}

The contribution to the energy of the sphere from the work done against
the buoyancy is formally $\int^\infty_{z_c} \hat{\bf z} \cdot {\bf f}\,
dz_c$, where ${\bf f}$ denotes the {\it exact\/} buoyant force.  If we add
and substract ${\bf f}_<$ to ${\bf f}$,  perform the integral
$\int^\infty_{z_c} \hat{\bf z} \cdot {\bf f}_<\, dz_c$, and then take into
account the gravitational energy as in Eq.(\ref{E>}), we get
\begin{eqnarray}
{\cal E}_<(z_c)&=&{E z_c\over 4GM}  +{N\hbar\over 2880\pi
GM}\left[{2Rz_c\over z_c^2-R^2}-\ln\left(z_c+R\over
z_c-R\right)\right] - {\cal C}
\label{E<}
\\
{\cal C}&\equiv& \int^\infty_{z_c} \hat{\bf z} \cdot ({\bf f}_<-{\bf f})\,
dz_c
\label{C}
\end{eqnarray}
Because ${\bf f}_<\rightarrow {\bf f}$ in the NHR, ${\cal C}$ is actually
constant so long as $z_c$ is taken in the NHR.   

Because outside the NHR part of the the buoyancy is best
described by wave scattering, and should be weaker than when calculated
entirely from the fluid model, we may infer that ${\cal
C}>0$.  A fair approximation to ${\cal C}$ can be had by replacing ${\bf
f}$ in Eq.(\ref{C}) by ${\bf f_>}$ over the range of $z_c$ for which ${\bf
f_<}\geq{\bf f_>}$, and setting ${\bf f}={\bf f_<}$ in the domain where
${\bf f_<} < {\bf f_>}$.  By comparing Eqs.(\ref{farforce}) and
(\ref{nearforce}) numerically one finds that the transition occurs at
$z_c=1.19826R$ for $\xi=10$, with almost no change as
$\xi$ varies by factors of 10 either way.  Using these values a numerical
integration gives ${\cal C}=0.00661(N\hbar/8\pi GM)$, confirming our hunch
about the sign of ${\cal C}$, but also showing that in natural units ${\cal
C}$ is small.  Hence, the fact that ${\cal C}$ has only been approximated
should not upset the following arguments.

The floating point, determined by the minimum of ${\cal E}_<(z_c)$, is
unaffected by the value of ${\cal C}$; it now occurs at $z_c=\ell R$ with
\begin{equation}
\ell =\sqrt{1+\sigma}
\label{ell}
\end{equation}
and $\sigma$ still defined by Eq.(\ref{sigma}).   For self--consistency this
should fall within the NHR: $\ell<1.01$.  By Eq.(\ref{ell})
we thus require $\sigma<0.02$.  This condition is fulfilled by all
macroscopic or mesoscopic objects ($\hbar/E\ll R$) unless the number of
radiance species
$N$ is very large. Position (\ref{ell}) coincides with that at which 
the object
displaces its own energy's worth of radiation, as we saw in
Sec.\ref{energeticsFR}. Thus in the NHR the buoyancy is Archimedean, and step 2
of UW's argument is upheld. 

The growth in black hole entropy when the sphere is dropped from
the floating point, namely $T_0{}^{-1}\,{\cal E}_<(\sqrt{1+\sigma} R)$, is
\begin{equation}
\Delta S_{BH} = (N/90)\,\big[\sigma^{-2}(1+\sigma/2)\sqrt{1+\sigma}
 - h(\sigma)/4-0.595\big]
\label{dSB}
\end{equation}
The displaced entropy is given by Eq.(\ref{Sdisp}).  It is easy to see that
for $\sigma<0.02$, $\Delta S_{BH}>{\cal S}_{\rm disp}$.  We
thus confirm step 1 of UW's argument.

Step 3 in UW's procedure maintains that ${\cal S}_{\rm disp} > S$
on the grounds that unconfined thermal radiation is  the most entropic form
of matter for given volume and energy. This generic entropy bound has been
questioned\cite{Bound3} because the two entities compared here, unconfined
thermal radiation, and matter confined to the object, are subject to
different boundary conditions.  Although violations of ${\cal S}_{\rm disp}
> S$ are rampant for systems whose various dimensions are very different,
they have not been found for spherical ones.  Accepting the inequality in
the spirit of this paper's scope, we find with UW that if the floating point
lies in the NHR, $\Delta S_{BH}>S$.  This, of course, guarantees that the
GSL will work as the spherical object falls in.

In the above argument it is the quantum buoyancy which protects the GSL. 
However, it does not follow, as UW\cite{UW1} would have it, that one
cannot derive the entropy bound (\ref{bound}) from the GSL, nor as
PW\cite{PW} would have it, that the entropy bound holds only
because of additional assumptions about the magnitude of $N$ and
$\hbar(RE)^{-1}$.   In this section we only assume that $\sigma\ll 1$, and
this in order that $\ell-1\ll 1$, so that the fluid method is accurate.

Let us now replace the $N$ in Eq.(\ref{dSB}) in terms of $\sigma^2$
according to Eq.(\ref{sigma}).  We find
\begin{equation}
\Delta S_{BH} = (2\pi RE/\hbar)\,\big[(1+\sigma/2)\sqrt{1+\sigma}-
(1/4)h(\sigma)\,\sigma^2-0.595\sigma^2\big]
\label{temp}
\end{equation}
Since the last two terms in the square brackets are negative, the GSL will
fail upon infall of the object unless at least
\begin{equation}
S<(2\pi RE/\hbar)\,(1+\sigma/2)\sqrt{1+\sigma}
\label{final}
\end{equation}
For the small $\sigma$ we are assuming this is equivalent to the entropy
bound (\ref{bound}), which is thereby seen to be a {\it necessary\/}
condition for the GSL.   We further discuss this derivation of the bound in
Sec.\ref{intermediate}.  Since it is carried out in the face of buoyancy, it
serves as justification for several recent extensions of the bound to
spinning\cite{spinning} and charged\cite{charged} entropy--bearing objects
whose derivations ignored buoyancy.  

We also note that the function of $\sigma$ in the square brackets in
Eq.(\ref{temp})  is positive for $\sigma < 0.02$.  Thus the entropy bound
(\ref{bound}) guarantees that $\Delta S_{BH}>S$: when the floating point
is in the NHR, the entropy bound is also a {\it sufficient\/} condition for
operation of the GSL. 

\section{The intermediate region}
\label{intermediate}

In the intermediate region
$1.01R{\,\,\lower0.5ex\hbox{$\sim$}\kern-10pt\raise0.5ex\hbox{$<$}\,\,}
z_c{\,\,\lower0.5ex\hbox{$\sim$}\kern-10pt\raise0.5ex\hbox{$<$}\,\,}2R$
neither of the previous calculations of the buoyant force is expected to be
accurate.  To calculate the buoyant force directly from scattering without
benefit of the approximationS $\lambda\gg R$ or $\lambda\ll R$ would be
very difficult.  Thus we propose to substitute ${\bf f}$ in the
intermediate region by an interpolation between ${\bf f}_>$ and ${\bf
f}_<$.  Specifically, we interpolate harmonically between the two
versions of the buoyancy energy, to wit
\begin{equation}
{\cal E}({z_c})={E\,z_c\over 4GM}+{\tilde{\cal E}_<({z_c})\ \tilde{\cal
E}_>({z_c})\over\tilde{\cal E}_<({z_c}) +\tilde{\cal E}_>({z_c})};\qquad
\tilde{\cal E}_j\equiv {\cal E}_j - {E\,z_c\over 4GM}  
\label{interpolate}
\end{equation}
The ${\cal E}_j$'s here are as defined by Eqs.(\ref{E>}) and (\ref{E<}),
the last with ${\cal C}$ set to zero since we found it to be relatively
small.  This interpolation for ${\cal E}$ is completely smooth, and obviously
has the correct asymptotics for $z_c-R\ll R$ and $z_c\gg R$.  Numerically we
find that
$-\partial{\cal E}/\partial z_c$ agrees with ${\bf f}_<$ to at least
$0.001\%$ for
$R<z_c<1.01R$, and with ${\bf f}_>$ to at least $0.05\%$ for $z_c>2R$.
Eq.(\ref{interpolate}) should thus be a reasonable good approximation 
to  ${\cal E}$ in the intermediate region. 

The interpolated ${\cal E}$ can be cast into the form
\begin{eqnarray}
{\cal E}(z_c)&=&ER(4GM)^{-1}\,\,{\cal F}(\xi,\sigma^2,z_c/R)
\label{Efinal}
\\
{\cal F}(\xi,\sigma^2,\upsilon)&\equiv& \upsilon+{\xi\sigma^2
(\upsilon^5+\upsilon^3)g(\upsilon)\over
4(\upsilon^2-1)\left[\xi (\upsilon^5+\upsilon^3)+28\pi^5 (\upsilon^2-1)^5\, 
g(\upsilon)\right] }
\\
g(\upsilon)&\equiv& 2\upsilon-(\upsilon^2-1)\ln\left({\upsilon+1\over 
\upsilon-1}\right)
\end{eqnarray}
${\cal F}$ as a function of $\upsilon=z_c/R$ has a single minimum in the
physical range
$\upsilon>1$.  Thus ${\cal E}$ is minimal at a value of $z_c/R$ - we
again denote it $\ell$ - which depends only on $\sigma$ and $\xi$.   Taking
$\xi=10$ we find numerically that $\ell$ increases monotonically with 
$\sigma$. 
In particular, it grows from 1.01 to 2 as $\sigma$ varies from 0.02 to
$105$.  In light of previous findings, the corresponding
range of $z_c$ bridges the gap between NHR and FR.

Replacing $ER$ in Eq.(\ref{Efinal}) by means of Eq.(\ref{sigma}) and
dividing by $T_0$ gives the minimum increase in black hole entropy:
\begin{equation}
\Delta S_{BH} ={N\over
90\sigma^2}\,\,{\cal F}\big(\xi,\sigma^2,\ell(\xi,\sigma)\big)
\label{DSBHfinal}
\end{equation}
This has to be compared with $S_{\rm disp}$ of Eq.(\ref{Sdisp}). 
Numerically we find (again for $\xi=10$) that $\Delta S_{BH}\leq S_{\rm
disp}$ for $\sigma\geq 1.42$, which contradicts step 1 of UW's argument.  The
transition point $\Delta S_{BH}=S_{\rm disp}$ occurs here at
$\ell=1.275$, a position close to the $\ell=1.24$ that we computed for the  
switch in Sec.\ref{energeticsFR} using FR formulae.  As already
discussed there,  under the circumstances one cannot naturally redesign
the buoyancy argument to understand why the GSL works.  That argument is of
no help when $N$ is large enough to make $\sigma$ somewhat larger than unity.

How does the entropy bound (\ref{bound}) fare ?  Evaluating $\sigma$ in 
the prefactor of Eq.(\ref{DSBHfinal}) we get
\begin{equation}
\Delta S_{BH} =(2\pi
ER/\hbar)\,\,{\cal F}\big(\xi,\sigma^2,\ell(\xi,\sigma)\big)
\label{DSBHx}
\end{equation}
Numerically ${\cal F}\big(\xi,\sigma^2,\ell(\xi,\sigma)\big)>1$ for $\ell>1$.
Hence, also in the intermediate $\sigma$ regime
($1.01{\,\,\lower0.5ex\hbox{$\sim$}\kern-10pt\raise0.5ex\hbox{$<$}\,\,} \ell
{\,\,\lower0.5ex\hbox{$\sim$}\kern-10pt\raise0.5ex\hbox{$<$}\,\,}  2)$, 
bound (\ref{bound}) implies the inequality $\Delta S_{BH}>S$, and is thus a
{\it sufficient\/} condition for the operation of the GSL when the object is
dropped in. 

To what extent is the entropy bound also a necessary condition for the
GSL: can the bound be derived from the GSL ?  Numerically we find ${\cal
F}\big(\xi,\sigma^2,\ell(\xi,\sigma)\big)<1.1$ for
$\sigma<0.1$.   Applying the GSL to our gedanken experiment allows us to
derive bound (\ref{bound}) for $\sigma<0.1$, apart from a
$<10\%$ correction.  We may reach the same conclusion from Eq.(\ref{final}). 
Bound (\ref{bound}) can also be derived for arbitrary $N$, apart from an
overall constant of crudely order unity, by considering free fall of the
object into a black hole\cite{MG8}.  Since in all cases where $S$ can be
computed explicitly for  non--black hole systems\cite{BekSchiff}, one finds
the true $S$ to always fall at least an order of magnitude below bound
(\ref{bound}), there is no point in introducing corrections to the
canonical form (\ref{bound}) on the basis of the above comments.

\section{Buoyancy of a charged elementary particle}
\label{elementary} 

All our previous considerations referred to macroscopic spherical objects. 
For $N$ not large these have very small $\sigma$, so that only the
NHR case is important (if we exclude the issue of calculating the offset
${\cal C}$).  It is, however, interesting to enquire into buoyancy of an
elementary particle, one whose radius is of order of its Compton length,
$\hbar/E$. The particle can no longer be regarded as a sphere with
reflecting walls; wave scattering from it must be described by a coupling of
the waves with the particle's interaction Hamiltonian.  This is obviously
detail dependent, and complicated in general.  But in the special case of
a structureless electrically charged particle, the (Compton) scattering is
well understood.  

The characteristic length associated with a singly charged particle is the 
Thomson radius $R_{\rm T}=e^2/E$.  Because $\alpha\equiv e^2/\hbar\approx
1/137$, $R_{\rm T}\ll \hbar/E$, so if it were possible to lower the particle
down to the horizon, its center's position $z_c\approx \hbar/E$ would still
be a few lengths  $R_{\rm T}$ away from it.  In this sense the
particle is always in the FR.   But, as we shall see, the buoyant
force is of NHR form.    The point is that for wavelengths $\lambda\gg
\hbar/E$, and averaged over the two polarizations, the scattering
crossection is
\begin{equation}
d\sigma/d\Omega = (R_{\rm
T}^2/2)\,\big[1+(\hat{\bf n}\cdot\hat{\bf n}')^2\big],
\label{Thomson}
\end{equation}
where $\hat{\bf n}$ and $\hat{\bf n}'$ are the incidence and scattering
directions, respectively.  This form is missing the $(R/\lambda)^4$ of
Eq.(\ref{crossection}); this explains why FR behavior does not set in.

Instead of Eq.(\ref{dP}) we have
\begin{eqnarray}
d{\bf P}/dt&=&(R_{\rm T}{}^2/2)\,\int d\lambda \int d\hat{\bf n}\,\hat {\bf
n}\,I(\lambda, \hat {\bf n})\,\int d\hat{\bf n}'\, [1+(\hat{\bf
n}\cdot\hat{\bf n}')^2]\,(1-\hat{\bf n}\cdot\hat{\bf n}')\\
&=&(8\pi R_{\rm T}{}^2/3)\,\int d\lambda \int d\hat{\bf n}\,\hat {\bf
n}\,I(\lambda, \hat {\bf n}).
\end{eqnarray}
Substituting Eq.(\ref{Planckian}) for $I$, doing first the $\lambda$ integral  
by rescaling and use of the identity $\int_0^\infty
x^3(e^x-1)^{-1}dx=\pi^4/15$, and identifying $t$ with the proper time $\tau$
in the orthonormal frame leads us to ($N=1$ here by assumption: the particle
interacts only with electromagnetic waves)
\begin{equation}
d{\bf P}/d\tau=(32\pi^6\hbar R_{\rm T}{}^2/45)\,\int d\lambda \int d\hat{\bf
n}\,\hat {\bf n}\,(T/2\pi\hbar)^4.
\label{newdP}
\end{equation}
As explained at the end of Sec.\ref{momentumcurved}, one must take $T$
as anisotropic, depending on the direction of incidence; since $R_{\rm T}$
is the characteristic radius of scattering here, it replaces $R$ in the
expression for $T$, Eq.(\ref{anis}), which we substitute in
Eq.(\ref{newdP}).  The integral over $\hat{\bf n}$ is performed as in
Sec.\ref{momentumcurved}.  Multiplying the result by
$d\tau/dt=\chi\approx z_c(4GM)^{-1}$ we get the buoyant force as measured at
infinity
\begin{equation}
{\bf f}={\hat{\bf z}\, \hbar \over
270\,\pi\,GM}\,{R_{\rm T}{}^3 \, z_c^2\over
\left(z_c^2-R_{\rm T}{}^2\right)^{3}}.
\label{elemforce}
\end{equation}
Because $z_c
{\,\,\lower0.5ex\hbox{$\sim$}\kern-10pt\raise0.5ex\hbox{$>$}\,\,}
\hbar/E \gg R_{\rm T}$, this force is of the same form as 
Eq.(\ref{nearforce}) for a macroscopic sphere in the NHR, if we replace
$R\rightarrow R_{\rm T}$.

Several caveats about this derivation are in order.  We have neglected
quantum corrections to the Thomson crossection (\ref{Thomson}) which are
given by the Klein--Nishina formula.  These corrections become important when
 $\lambda {\,\,\lower0.5ex\hbox{$\sim$}\kern-10pt\raise0.5ex\hbox{$<$}\,\,}
2\pi\hbar/E$. As we saw in Sec.\ref{new}, $\bar\lambda\sim 10z$, and so as
the particle approaches the horizon ($z_c \sim \hbar/E$), neglect of these
corrections is no longer entirely justified.  However, further out
Eq.(\ref{elemforce})  should be accurate in this respect.  We have also
neglected quantum corrections arising from the fact that by the uncertainty
principle, the particle cannot be at rest and well localized at once, as we
assume implicitly when we appeal to buoyancy.  Further, the force
(\ref{elemforce}) can only represent an average: the particle should be
subject to buffeting originating in thermal fluctuations of the Unruh
radiance. 

Because $z_c\gg R_{\rm T}=\alpha(\hbar/E)$, we can
recast Eq.(\ref{elemforce}) into the form
\begin{equation}
{\bf f}\approx (2/135\pi)\alpha^3 (\hbar/E z)^4\, {\bf f}_{\rm grav},
\end{equation}
where ${\bf f}_{\rm grav}=\hat{\bf z}\,E\,(4GM)^{-1}$ is the gravitational
force as measured at infinity.  Obviously, $|{\bf f}|\ll |{\bf f}_{\rm
grav}|$ over the physical range $z_c>\hbar/E$.  Hence buoyancy is negligible
for our particle; there is no floating point, and buoyancy corrections to the
energetics are entirely negligible.  The neglected quantum corrections are
unlikely to change this extreme state of affairs.  In recent arguments
supporting the uniformly spaced black hole area
quantum spectrum\cite{MG8,Brazil,Hod}, the buoyancy of an elementary
particle was ignored.  The present work supplies a basis for that neglect.

\section{DISCUSSION}
\label{discussion} 

The effects of quantum buoyancy are parametrized by $\sigma$ defined in
Eq.(\ref{sigma}).  In a world with few radiation species,
$\sigma\approx 10^{-2}, 10^{-5}$, and $10^{-18}$ for an average nucleus, an
average atom, and a grain of salt, respectively.    All these objects would
thus have floating points almost at the horizon ($z_c-R\ll R$).  The results
of Secs.\ref{energeticsNHR} and \ref{intermediate} then tell us that buoyancy
protects the GSL upon infall of these or larger objects into a black hole. 
They also allow us to derive the universal entropy bound (\ref{bound}) from
the GSL, apart from a tiny correction which can probably be neglected.  For
macroscopic and mesoscopic objects, the entropy bound is thus a
{\it necessary\/} condition for the operation of the GSL, provided the
number of radiant species $N$ is not very large. 

As the number of radiation species $N$ is increased, $\sigma$ scales as
$\surd N$, and will eventually get to be a few tenths of unity for any
specified object.   The entropy bound for that object cannot then be derived
from the GSL as done previously.  It stops being a necessary
condition for the GSL, but remains a sufficient condition for the law to
function, this for arbitrary $N$ or $\sigma$.  Although it seems highly
unlikely that the world's particle spectrum contains very many species, UW
and PW have urged to consider what keeps the GSL working in such a
hypothetical eventuality.  Their suggestion that buoyancy does the job for
any $N$ is overturned by our finding  (Sec.\ref{intermediate}) that for
$\sigma$ exceeding 1.42 ($N>10^{9}$ for an atomic sized system), buoyancy
becomes insufficient to protect the GSL.  

The fact that the universal entropy bound remains a sufficient
condition for the GSL when $N$ gets large motivates us to consider the bound
as the more fundamental principle.  This is in the spirit of
today's outlook that holographic--type principles are very deep.
For weakly gravitating systems the entropy bound implies the standard
holographic principle\cite{MG7,Bousso}.  Neither entropy bound nor
holographic principle refers to the number of particle species in nature. 
One might worry that this is wrong: if one has more species to split a fixed
energy among, should not the possible entropy be higher the higher $N$ ? 
But if this were true, the arguments at the end of Sec.\ref{energeticsFR} and
at the end of Sec.\ref{intermediate} showing that the entropy bound
guarantees that the generalized entropy grows upon accretion of the object by
a black hole could fail.

We thus retain the standard form (\ref{bound}) of the bound.  In our
opinion, the absence of $N$ from the bound, from the holographic principle,
and from the formula for black hole entropy, can have two possible
origins.  One is that the deep logic of physical theory requires a
specific particle spectrum and a definite - not too large - $N$ for which
the black hole entropy attains its accepted value, and for which entropy
bound and holographic principle are still respected, growth of entropy with
$N$ notwithstanding.  The second possibility is that a proliferation of
particle species would entail interactions between them to such an
extent that it would change the way in which entropy scales up with $N$
for noninteracting species, causing it to saturate at large $N$ at a
value consistent with entropy bound and holographic principle.

\acknowledgments
This research is supported by a grant from the Israel Science Foundation,
established by the Israel Academy of Sciences and Humanities.

\end{document}